\documentclass[fleqn,usenatbib]{mnras}
\usepackage{newtxtext,newtxmath}
\usepackage[T1]{fontenc}

\usepackage{graphicx}	
\usepackage{amsmath}	
\usepackage{color}

\def\wisk#1{\ifmmode{#1}\else{$#1$}\fi}
\def\wm2sr {Wm$^{-2}$sr$^{-1}$ }		
\def\nw2m4sr2 {nW$^2$m$^{-4}$sr$^{-2}$\ }		
\def\nwm2sr {nWm$^{-2}$sr$^{-1}$\ }		
\def\nw2m4sr {nW$^2$m$^{-4}$sr$^{-1}$\ }

\def\lt     {\wisk{<}}

\def\le     {\wisk{_<\atop^=}}

\def\lsim   {\wisk{_<\atop^{\sim}}}
\def\gsim   {\wisk{_>\atop^{\sim}}}

\def\mic {\wisk{ \mu{\rm m }}}

\title[Short title, max. 45 characters]{MNRAS \LaTeXe\ template -- title goes here}

\title[Probing the rest-frame of the Universe]{Probing
the rest-frame of the Universe with near-IR cosmic infrared background}

\author[A. Kashlinsky \& F. Atrio-Barandela]{
A. Kashlinsky$^{1}$
\thanks{E-mail: Alexander.Kashlinsky@nasa.gov}
and
F. Atrio-Barandela$^{2}$\thanks{E-mail: atrio@usal.es}
\\
$^{1}$ Code 665, Observational Cosmology Lab, NASA Goddard Space Flight Center,
Greenbelt, MD 20771 and SSAI, Lanham, MD 20706\\
$^{2}$ Department of Fundamental Physics, University of Salamanca, 37008 Salamanca, Spain
}

\date{Accepted 22 April 2022. Received 13 April 2022; in original form 16 March 2022}

\pubyear{2022}

\begin{document}
\label{firstpage}
\pagerange{\pageref{firstpage}--\pageref{lastpage}}
\maketitle

\begin{abstract}
While the cosmic microwave background (CMB) dipole is largely assumed entirely kinematic, there appears evidence that a part of it is primordial. Such possibility arises in models implying a tilt, interpreted as a dark flow, across the observable Universe. 
The kinematic nature of the entire CMB dipole can be probed using the dipole of cosmic backgrounds from galaxies after the last scattering. 
The near-IR cosmic infrared background (CIB) spectral energy distribution leads to an amplified dipole compared to the CMB. The CIB dipole is affected by galaxy clustering, decreasing with fainter, more distant galaxies, and by Solar System emissions and Galactic dust, which dominate the net CIB cosmological dipole in the optical/near-IR. We propose a technique that enables an accurate measurement of the kinematic near-IR CIB dipole. The CIB, effectively the integrated galaxy light (IGL), would be reconstructed from resolved galaxies in the forthcoming space-borne wide surveys covering four bands 0.9 to 2.5 \mic. The galaxies will be sub-selected from the identified magnitude range where the dipole component from galaxy clustering is below the expected kinematic dipole. Using this technique the dipole can be measured in each of the bands at the statistical signal-to-noise $S/N\gsim$50--100 with the forthcoming {\it Euclid} and {\it Roman} surveys, isolating CMB dipole's kinematic nature.
\end{abstract}




\textcolor{black}{The cosmic microwave background (CMB) dipole is $\delta T_{\rm CMB, dip}=3.346$mK toward $(l,b)_{\rm Gal} = (263.85, 48.25)^\circ$ measured with signal-to-noise ratio $(S/N)\gsim200$ 
\citep{Kogut:1993,Fixsen:1994}}. 
Its amplitude $D_{\rm CMB}\equiv \delta T_{\rm CMB, dip}/T_{\rm CMB}=1.25\times 10^{-3}$ is  mostly, but not unanimously \citep{Gunn:1988}, interpreted as Sun's motion at $V_{\rm CMB}=370$ km/sec if CMB \textcolor{black}{traces the Universe's rest-frame commonly identified as the universal expansion's rest-frame}. 

There appeared assertions contradicting CMB dipole's kinematic interpretation. Analysis of the cumulative kinematic Sunyaev-Zeldovich dipole at CMB locations from an all-sky X-ray cluster sample suggested what was termed the dark flow of the clusters with respect to CMB extending to $\sim 1$Gpc, the signal persisting in WMAP 3,5,7,9 yr CMB data \citep{Kashlinsky:2008,Kashlinsky:2009,Kashlinsky:2010,Kashlinsky:2012} and Planck 1-yr data \citep{Atrio-Barandela:2013,Atrio-Barandela:2015}. Comparing gravity dipole with velocity provides similar tests \citep{Villumsen:1987} and a non-kinematic CMB dipole would be broadly consistent with the misalignment in direction and/or amplitude of the  gravitational force reconstructed from galaxy and cluster catalogs from the CMB dipole \citep[e.g.][]{Gunn:1988,Lavaux:2010,Erdogdu:2006,Kocevski:2006,Wiltshire:2013}, the various claims of peculiar flows and their properties \citep[e.g.][]{Mathewson:1992,Lauer:1992,Ma:2011,Colin:2019}, the radio-counts dipole \citep{Nodland:1997,Jain:1999,Singal:2011}, the recent {\it WISE} source-counts dipole \citep{Secrest:2021} and the anisotropy in X-ray cluster scaling relations \citep{Migkas:2020}. \textcolor{black}{Such measurements achieved only a limited significance, $S/N\sim$(3-4), necessitating new ideas for reaching higher $S/N$ required to robustly probe the CMB dipole's kinematic nature}.

Before inflation was developed, it was proposed that CMB dipole may be primordial, not entirely kinematic \citep{King:1973,Matzner:1980}.
Cosmological (inflation-produced) curvature perturbations have zero intrinsic dipole at last scattering \citep{Turner:1991}, so primordial CMB dipole would reflect pre-inflationary conditions from tilt \citep{King:1973} by space inhomogeneities before inflation \citep{Turner:1991} pushed beyond the cosmological horizon as constrained by the CMB quadrupole anisotropy \textcolor{black}{\citep{Kashlinsky:1994,Turner:1991,Grishchuk:1992,Das:2021,Tiwari:2022}}, or could arise from entanglement of our Universe with super-horizon domains in certain Multiverse models \citep{Mersini-Houghton:2009}. 

The  CMB dipole nature can be probed with dipoles of cosmic backgrounds (CBs)  from galaxies after the last scattering, which reflect the universal expansion rest-frame. If the Sun moves at $V\ll c$ relative to distant sources producing the CB its intensity $I_\nu$ at frequency $\nu$ would have dipole in Sun's rest-frame known as the Compton-Getting effect for cosmic rays \citep[e.g.][]{Gleeson:1968}:
\begin{equation}
\boldsymbol{d}_\nu= (3-\alpha_{\nu,\infty}) \frac{\boldsymbol{V}}{c} \bar{I}_\nu
\label{eq:dipole_alpha}
\end{equation}
where $\alpha_{\nu,\infty}=\partial{\ln I_\nu}/\partial{\ln \nu}$ and the subscript $\infty$ indicates $I_\nu$ from integrating over the entire range of fluxes/magnitudes of the \textcolor{black}{sources contributing to the given band}. Eq. \ref{eq:dipole_alpha} follows since the Lorentz transformation leaves $I_\nu/\nu^3$ invariant \citep{Peebles:1968}.
If CMB traces the Universe's rest-frame, $\boldsymbol{V}=\boldsymbol{V}_{\rm CMB}$ for any CB. Various wavelength CBs exhibit $\alpha_{\nu,\infty}<2$ such that their dipole amplitude gets {\it amplified} compared to the CMB dipole, 
if purely kinematic: such wavelengths include X-rays \citep{Fabian:1979}, radio \citep{Ellis:1984,Itoh:2010}, and sub-mm \citep{Kashlinsky:2005,Fixsen:2011}. 
An all-sky CB dipole measured with signal/noise $S/N$ will have its direction probed with the directional accuracy of $\Delta\Theta_{\rm dipole}\simeq \sqrt{2}(S/N)^{-1}$radian \citep{Fixsen:2011}.

 
The near-IR CIB has the well-defined $\alpha_{\nu,\infty}\ll 2$ from $< 1\mic$ to $\simeq 4\mic$ and, with enough sky, its (amplified) dipole can be probed to resolve deviations from the CMB dipole in amplitude and direction, provided contributions from diffuse Galactic (cirrus) and Solar System (zodi) foregrounds, and from galaxy clustering are negligible. In this {\it Letter} we demonstrate that the foreground contributions dominate the cosmological dipole for the {\it net} near-IR CIB and propose a method to eliminate them with the integrated galaxy light (IGL) part of the CIB by using resolved galaxies. This technique is applicable to any
  sufficiently wide-field sky survey in which individual galaxies can
  be identified and which goes deep enough to minimize the dipole due
  to large-scale structure. Such configurations may be available with the cosmic infrared background (CIB) from the forthcoming space missions {\it Roman} \citep{Spergel:2015} and {\it Euclid} \citep{Laureijs:2011}, which has an ongoing near-IR CIB program 
 \citep{LIBRAE:2013,Kashlinsky:2018}. 
 We then identify the CIB configuration for a highly precise measurement of the CIB dipole in these surveys to resolve the kinematic nature of the long-known CMB dipole and
discuss the various other components of the proposed measurement and how the near-IR CIB dipole is spectrally distinguished from any remaining foreground contribution. 


\section{Net near-IR CIB dipole vs foregrounds}

\begin{figure}
\includegraphics[width=3.25in]{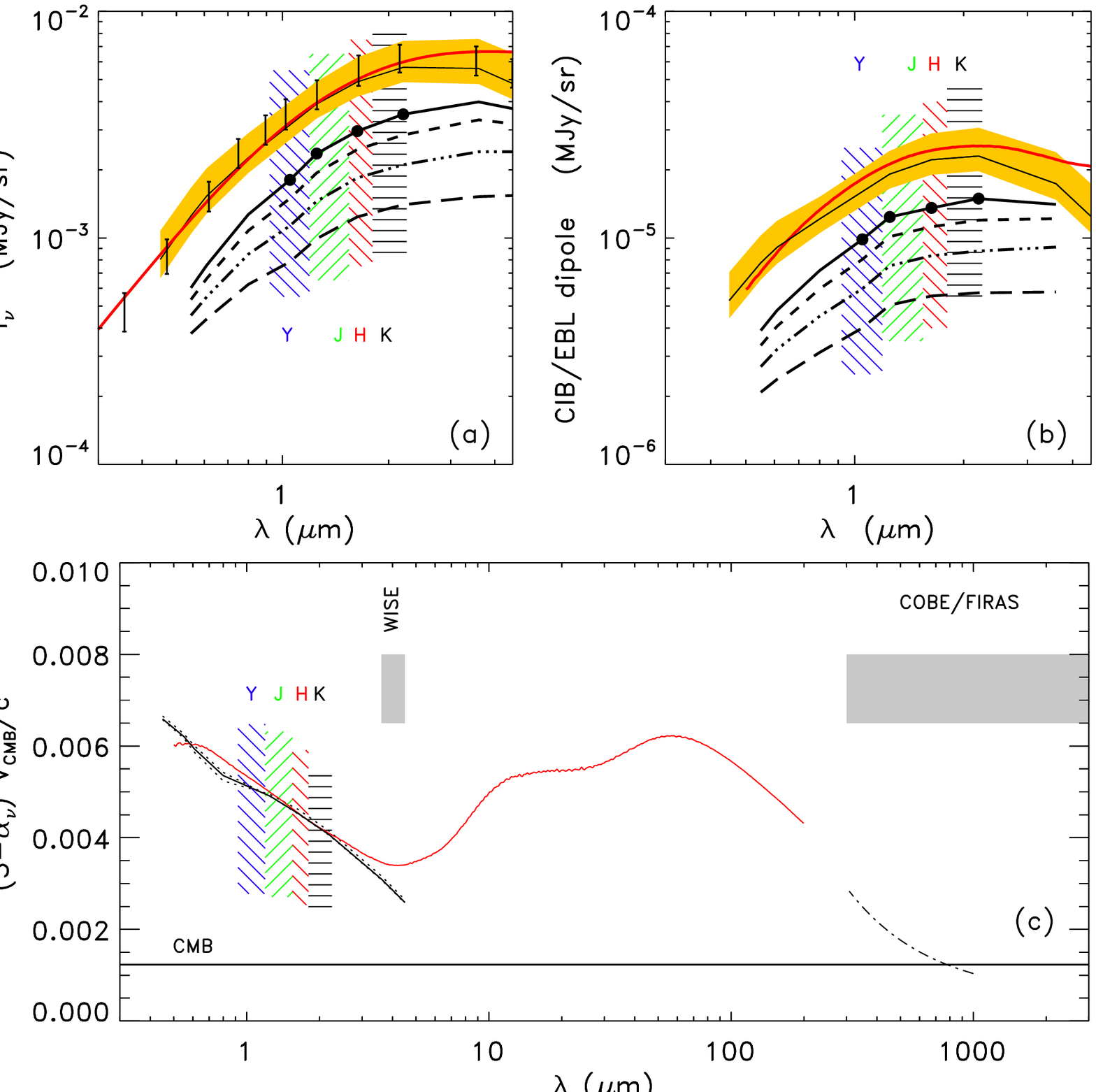}
\caption{\small (a) CIB $I_\nu$: black errors show CIB by the net observed deep counts, 
yellow is the HRK reconstruction \citep{Helgason:2012} with the solid black line showing 
the default model.  Red line is from \citet{Driver:2016}. Black thick lines 
cover CIB from galaxies over $m_0<m<m_1$ with $m_0=$18,19,20,21 (solid, dashed, 
dashed-triple-dotted,  long-dashed). Filled circles are for 
galaxies over $18\leq m_{\rm AB}\leq 24$. (b) CIB dipole if the entire CMB 
dipole is kinematic.  The near-IR filters are marked.
(c) CB dipole amplitude from optical to sub-mm shows the advantage of the near-IR 
configuration over previous studies employing sub-mm COBE/FIRAS low angular resolution data 
\citep[][]{Fixsen:2011} 
and mid-IR WISE AGN catalog \citep{Secrest:2021}. Black lines are for the HRK-reconstructed CIB 
(dotted lines show the HFE$\rightarrow$LFE spread), red line is the CIB/EBL from 
\citet{Driver:2016}, the black dashed-dotted line corresponds to the sub-mm CB amplification
\citep[][]{Kashlinsky:2005}; CMB dipole is marked.}
\label{fig:cib}
\end{figure}

The net CIB is made up of unresolved CIB and the part due to resolved galaxies. Fig.\ref{fig:cib}a shows $I_\nu$ integrated over known galaxies from counts data; its contribution peaks at $m_{\rm AB}\!\sim$20 \cite[Fig. 1 in][and below]{Kashlinsky:2019}. 
Red line shows the fit using observed galaxy spectra \citep{Windhorst:2011,Driver:2016}. Yellow-filled regions show the empirical CIB reconstruction from the multi-epoch, multi-wavelength known galaxy luminosity function data \citep[][hereafter HRK]{Helgason:2012}.
Black lines show the HRK-reconstructed background ("default", the optimal value); the yellow shaded region marks the high-faint-end (HFE) and low-faint-end (LFE) limits of the observed luminosity function data used by HRK. 
The contribution from new sources implied by the source-subtracted CIB fluctuations uncovered in {\it Spitzer} measurements \citep{Kashlinsky:2005a,Kashlinsky:2007,Kashlinsky:2012a} contributes little to the overall mean CIB \citep{Kashlinsky:2007a} and will not enter the final configuration proposed below \citep[see review][]{Kashlinsky:2018}. The Y, J, H, K bands displayed approximate the bands available on {\it Euclid} and {\it Roman}.

Fig.\ref{fig:cib}b shows the expected CIB dipole assuming 1) the entire CMB dipole is of kinematic origin, and 2) no excess in CIB levels above that from known galaxies \citep{Kashlinsky:2018}.  If CMB \textcolor{black}{traces} the Universe's rest-frame the CIB dipole should point in the CMB dipole direction and have the corresponding amplitude, eq.\ref{eq:dipole_alpha}.  

Fig.\ref{fig:cib}c compares the relative CB dipole 
at the near-IR bands with previous probes. At the near-IR bands the CB dipole  should be amplified over that of the CMB to $\left(d_\nu/I_\nu\right) \simeq$(3.1--5.5)$D_{\rm CMB} \simeq$(4--7) $\times 10^{-3}$. 
The higher amplification coupled with the wide sky coverage will result in the high precision CIB dipole measurement if foregrounds can be eliminated as discussed below. 

\begin{figure}
\hspace*{-0.1in}\includegraphics[width=3.5in]{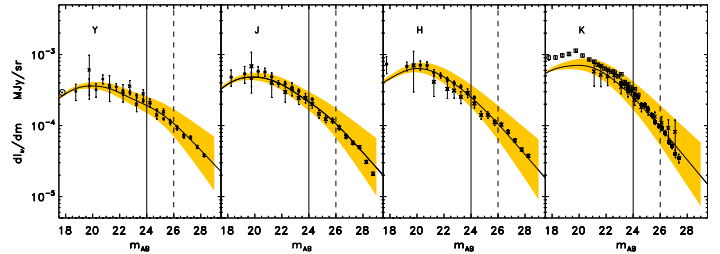}
\caption{\small CIB from known galaxy populations. Data at Y,J,H are from \citep[][and refs therein]{Windhorst:2011} and at K from \citep[][]{Maihara:2001,Keenan:2010}.  Solid lines show the default HRK reconstruction and the shaded areas mark the HFE$\rightarrow$LFE limits. Vertical solid and dashed lines mark {\it Euclid}'s Wide Survey's  and {\it Roman} 's magnitude limits. The measured K counts are taken as proxy for the {\it Roman} F184 band.}
\label{fig:cib-counts}
\end{figure}

Fig.\ref{fig:cib-counts} shows the CIB build-up at the near-IR bands. The HRK-reconstructed CIB fits well the observed data; we use the reconstruction for estimates below\textcolor{black}{ , but when the data from the new forthcoming missions becomes available it will be used directly in the analysis.}  

\begin{figure}
\includegraphics[width=2.75in]{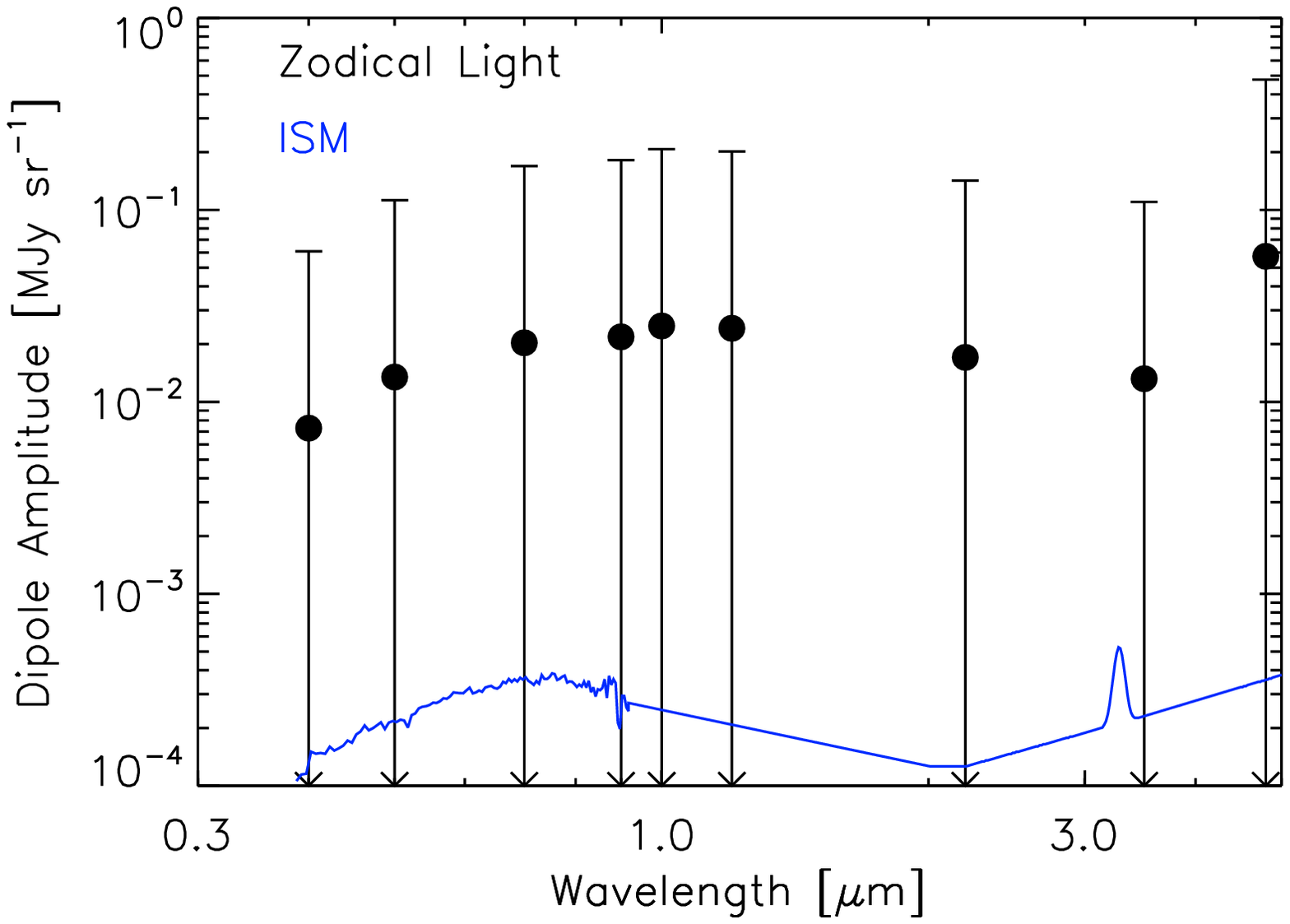}
\includegraphics[width=2.75in]{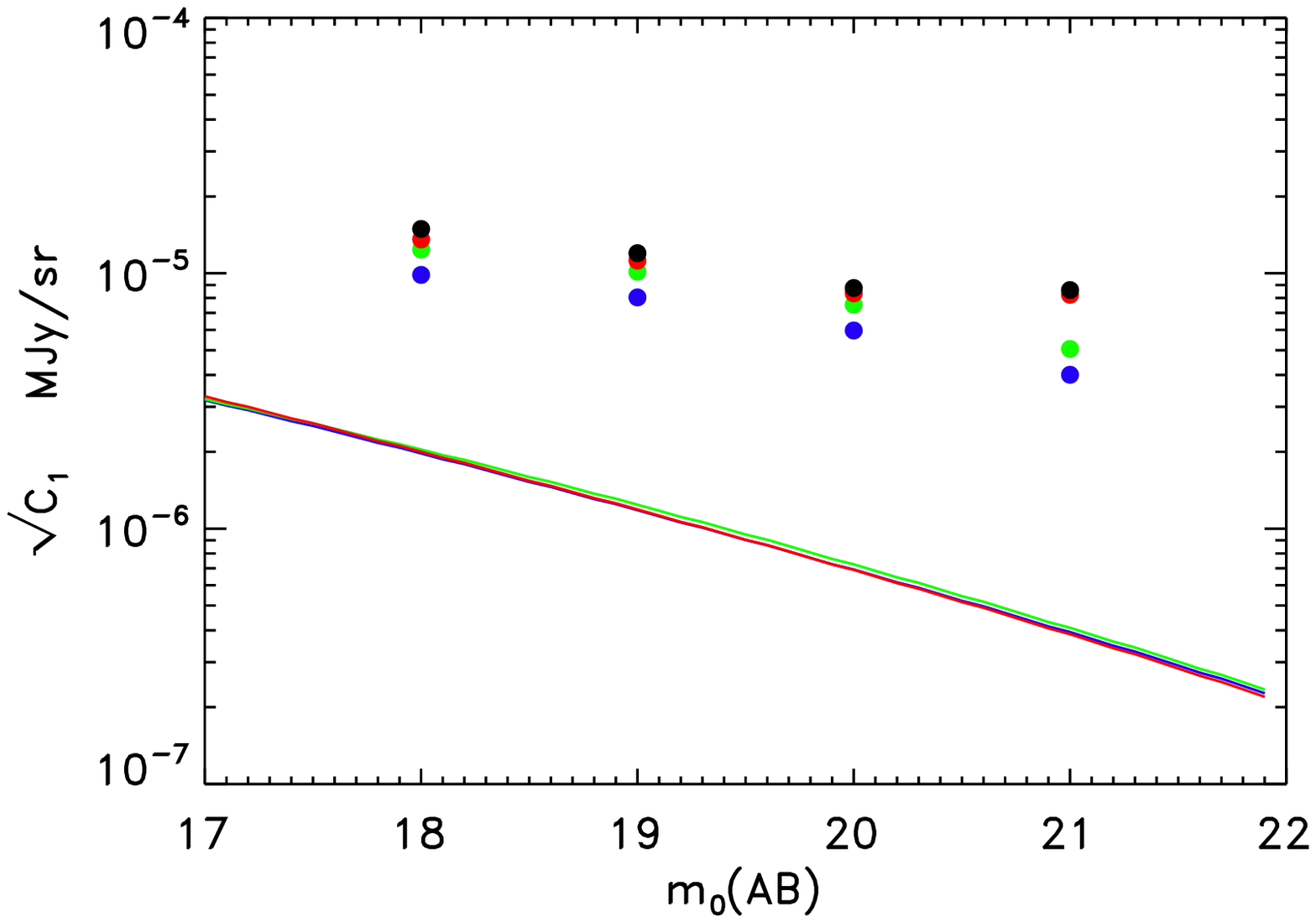}
\caption{\small 
{\bf Top}:  Zodiacal light dipole in near-IR bands could be $\sim0.1$ MJy sr$^{-1}$ if
measured at elongations $\sim90^\circ\pm10^\circ$ on a single
day (black upper limits). When averaged over long periods ($>1$ year), the net dipole amplitude
should decrease by an order of magnitude (black dots). The dipole amplitude
from scattered light and emission by the diffuse ISM (blue line) from scaling
the measured 100 $\mu$m emission after masking 85\% of the sky such that the
measurements are from only the lowest brightness regions near the Galactic
poles. Subtracting models of the zodiacal light and ISM/cirrus-scattered light
may reduce these by over an order of magnitude.
 {\bf Bottom}:  
 The lines show the clustering dipole at $m_1=24$ vs $m_0$ with the color notation for the Y,J,H,K bands from Fig. \ref{fig:cib};. Blue, green, red, black circles show the dipole expected at the given $m_0$ for Y, J, H , K bands. In the K (F184) band {\it Roman} will go 2 magnitudes deeper which will help reduce further the clustering dipole.
}
\label{fig:zodi}
\end{figure}
The Solar System and Galaxy foregrounds present obstacles to probing the kinematic 
dipole nature using the {\em net} near-IR CIB. Based on examination of compiled zodiacal 
light observations \cite[][]{Leinert:1998} and the DIRBE 100 $\mu$m images \citep[][]{Hauser:1998}, 
we estimate uncorrected near-IR dipoles in Fig.\ 3 (top)
for observations constrained to solar elongations near $90^\circ$ (as is common for space-based 
observations). The apparent dipole on a given date is $d_{\rm Zodi} \sim 0.11$ MJy sr$^{-1}$ at 1.25 $\mu$m. 
Since the direction of this dipole varies annually, averaging over one or more years reduces its net amplitude by $\sim10\times$. 
Even if the zodical light is modeled/subtracted with 1\% accuracy, the residual
dipole interferes with the CIB dipole. 
The cirrus dipole was estimated by simultaneous fitting for the 
monopole+dipole+quadrupole map components with several increasingly severe masks defined by brightness and 
Galactic and ecliptic latitude. 
The minimum dipole amplitude is $\sim$0.1 MJy sr$^{-1}$  at 100\mic\ 
 \citep[similar to][]{Fixsen:2011}. 
The dipole is oriented toward the Galactic center, even in the heavily masked cases.   We scale the 100 $\mu$m ISM dipole amplitude to 
the shorter wavelengths using the ISM colors from 
\cite{Arendt:1998,Sano:2016}. This yields a 1.25 $\mu$m ISM/cirrus dipole 
$2\times10^{-4}$ MJy sr$^{-1}$,  $\sim10\times$ 
greater than the expected CIB dipole.
To measure the CIB dipole would require
subtracting the ISM model with sub-percent accuracy. However, measurements of ISM colors typically show variation by factors of a 
few at different sky locations. 
Thus, while a simple scaling of the far-IR ISM emission can be used
to estimate the near-IR ISM dipole amplitude, it is unlikely to be sufficient here so {\it foregrounds present a substantial problem for probing the dipole of the net CIB}. The problem can be resolved with the method proposed next.

\section{IGL dipole from space-borne wide surveys}


To overcome the obstacles from foreground dipoles, we use the all-sky part of the CIB known as IGL (Integrated Galactic Light), reconstructed from resolved galaxies in a space-borne wide survey, 
\begin{equation}
I_\nu(l,b) = S_0\int_{m_0}^{m_1} 10^{-0.4m} \left[\frac{dN_\nu}{dm}\right] dm
\label{eq:dndm}
\end{equation}
 where $S_0=3.631\times10^{-3}$MJy, with $m_0$ suitably selected to remove the galaxy clustering dipole and $m_1 $ imposed by the wide 
survey sensitivity limits. Here we will want to measure the dipole in the flux, not the number counts. \textcolor{black}{For the dipole measurement the IGL will be constructed from the survey by adding fluxes from galaxies within the magnitude limit over each field-of-view (FOV) centered at $(l,b)$ with $>10^5$galaxies/deg$^2$ at $20<m_{\rm AB} <24$ \cite[e.g.][]{Windhorst:2011,Driver:2016}; it has similar spectral dependence to the CIB in Fig.\ref{fig:cib}.} 
To measure CIB dipole at the near-IR bands with $S/N$ significance requires $\simeq \left(S/N\right)\left(I_\nu/d_\nu\right)^2$ sources achieving $\left(S/N\right)\gsim (10-25)$ with $10^6$ independent positions if $d_\nu\sim(3-5)\times 10^{-3}I_\nu$ if foregrounds can be overcome. {\it Euclid} \citep{Laureijs:2011} will cover $\sim$15,000 deg$^2$ in Y, J, H bands to $m_{\rm AB}=24$ beyond the peak of their CIB contribution and {\it Roman} \citep[][]{Spergel:2015} is planned to cover $\sim 2,000$ deg$^2$ in Y, J, H and K (taken to aproximate F184 by {\it Roman}, \cite{Akeson:2019}) filters.

Foregrounds contribute negligibly to the IGL dipole, being removed locally in galaxy samples at an unprecedented precision \citep{Scaramella:2021}. Additionally, zodi is known to have no spatial fluctuations to very low levels \citep[e.g.][]{Arendt:2016} and Galactic cirrus has negligible fluctuation power \citep[][]{Gautier:1992} on individual galaxy scales \citep[see further refs in][]{Kashlinsky:2018}. 

The CIB dipole from galaxy clustering decreases with fainter galaxies \citep{Gibelyou:2012} and/or sources at higher $z$, constraining $m_0$. As part of their dark energy science, {\it Euclid} will measure spectroscopic redshifts over its extragalactic sky out to $z$=2.1 while {\it Roman} will go deeper.  The source redshift distribution is expected to have a mean at $z\sim 0.9$ with a density of 30 galaxies per square arcmin \citep{Laureijs:2011} and luminous giant galaxies can be detected out to $z\sim 3$; the linear scales subtended by the dipole at these epochs are well into the Harrison-Zeldovich regime. Fig.\ref{fig:zodi} bottom shows the remaining clustering dipole for galaxies with $m_0<m<m_1$ compared with the CIB dipole if the CMB dipole is kinematic. It was computed from the HRK reconstruction of CIB power with $m_1=24$ for Y/J/H/K which follows the Harrison-Zeldovich regime at large angular scales, i.e. $P=A(m_0)\ell$.  We then evaluate, with HRK reconstruction, $A(m_0)$  for given $m_0$  with the above $m_1$, and take the CIB dipole from clustering as $d_\nu=\sqrt{C_1}$ with the rms dipole $C_1=A(m_0)/\pi$. 
Additionally, cosmic variance affects the measured dipole, 
so at 95 (99) per cent confidence level $d_\nu<4  (5.7) \sqrt{C_1}$. \textcolor{black}{Selecting $m_0\sim$(18--21) leaves $\sqrt{C_1}$ over an order of magnitude below that expected from the kinematic CMB dipole}. 

That configuration, eq.\ref{eq:dndm}, has additional contribution to the dipole \textcolor{black}{due the local motion} $\Delta m=-(2.5\lg e) [1+\eta(m)]\frac{V}{c}\cos\Theta$ for CIB sources with SED $f_\nu\propto \nu^\eta$ \citep{Ellis:1984,Itoh:2010}.
In that case the dipole eq. \ref{eq:dipole_alpha} will be modified by the $m_{0,1}$ variation:
\begin{equation}
\boldsymbol{d}_\nu (m_0\!<\!m\!<\!m_1)= [(3-\alpha_{\nu,m_0\!<\!m\!<\!m_1})+\Delta \alpha_\nu]\frac{\boldsymbol{V}}{c} \bar{I}_\nu(m_0\!<\!m\!<\!m_1)
\label{eq:dipole_final}
\end{equation}
with
\begin{equation}
\Delta \alpha_\nu=\frac{\frac{dI_\nu}{dm}|_{m_1}}{\bar{I}_\nu(m_0\!<\!m\!<\!m_1)}[1+\eta(m_1)]-\frac{\frac{dI_\nu}{dm}|_{m_0}}{\bar{I}_\nu(m_0\!<\!m\!<\!m_1)}[1+\eta(m_0)]
\label{eq:dipole_correction}
\end{equation}
Figs.\ref{fig:cib},\ref{fig:cib-counts} show that at the values $m_{0,1}$ selected here over the near-IR bands the galaxy CIB contribution is only $\frac{dI_\nu}{dm}|_{m_{0,1}}\sim$ (a few percent)$\bar{I}_\nu(m_0\!\!<\!m\!\!<\!\!m_1)$. At these wavelengths one would probe galaxy stellar populations at various parts of the restframe spectrum, which is theoretically well studied and with effective $\eta\sim$0--2, depending on the age, mass-function, metallicity and epoch of the populations leading to $\Delta \alpha_\nu\ll \alpha_\nu$. When the measurement is performed, the values of $\eta$ and their distribution will be measurable for the cataloged galaxies and $\alpha, \Delta\alpha$ will be reconstructed directly. Fig. \ref{fig:cib} shows that the considered here CIB has a robust value of $\alpha$ at the near-IR wavelengths making the application of the CIB-based eq. \ref{eq:dipole_final} advantageous to achieving robust results with a high statistical accuracy. The statistical signal/noise for the dipole at each band is:
\begin{equation}
\left(\frac{S}{N}\right)_\nu=
127\;\frac{(3\!-\!\alpha_{\nu,m_0\!<\!m\!<\!m_1})D_{\rm CMB}}{4\times10^{-3}}\sqrt{\frac{N_{\rm tot}}{10^9}}
\label{eq:eq_s2n}
\end{equation}
Overall such $S/N$ can achieve sub-degree direction resolution.

\begin{figure}
\includegraphics[width=\columnwidth]{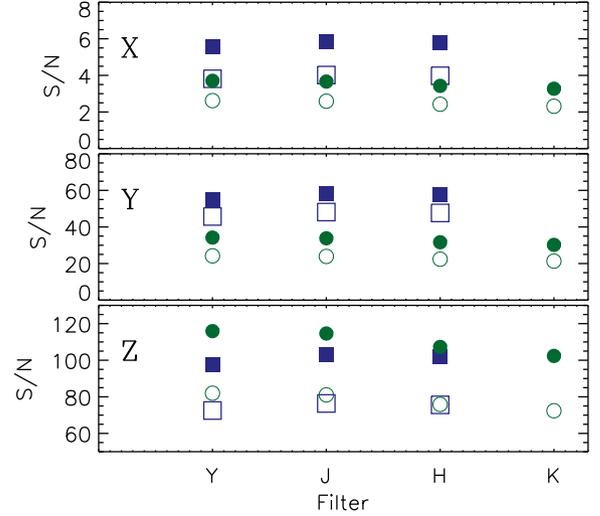}
\caption{\small $S/N$ for each of the {\it Euclid} (squares) and {\it Roman} (circles) photometric bands. Open symbols show 3 yrs of {\it Euclid} and the Southern patch of {\it Roman}. Filled symbols correspond to 6 yrs of {\it Euclid} and the Southern$+$Northern {\it Roman} patches. Other configurations 
would have $S/N\propto N_{\rm tot}^{1/2}$.}
\label{fig:cib-s2n}
\end{figure}

\begin{table}
\centering
\caption{HRK-reconstruced $N_{\rm tot}(m_0\!<\!m\!<\!m_1)/10^6$ per $1,000\;{\rm deg}^2$}
\begin{tabular} {|r|r|r|r|r|r|}
 \hline 
{ \footnotesize Filter} & {\footnotesize Y} & {\footnotesize J} & {\footnotesize H} & {\footnotesize K} \\
{ \footnotesize $m_0$} & { \footnotesize $m_1$=24/26} & {\footnotesize $m_1$=24/26} & { \footnotesize $m_1$=24/26} & { \footnotesize $m_1$=26} \\
 \hline
{ \footnotesize 18} & {\footnotesize 85.6/347.0} &  {\footnotesize 105.2/375.8} & {\footnotesize 128.8/411.3}  & {\footnotesize 422.7}  \\
{ \footnotesize 19} & {\footnotesize 84.9/346.2} &  {\footnotesize 104.3/372.2} & {\footnotesize 127.7/410.1} & {\footnotesize 421.2}   \\
{ \footnotesize 20} & {\footnotesize 82.9/344.2} &  {\footnotesize 101.6/372.7} & {\footnotesize 124.2/406.6} & {\footnotesize 417.0}   \\
{ \footnotesize 21} & {\footnotesize 78.0/339.4} &  {\footnotesize 95.2/365.7} & {\footnotesize 115.8/398.3} & {\footnotesize 407.2}  \\
\hline
\end{tabular}
\label{tab:tab_ngals}
\end{table}
Table \ref{tab:tab_ngals} 
shows the HRK-reconstructed number of galaxies, $N_{\rm tot}$,  expected in {\it Euclid} data with $m_0<m<m_1$ at $m_1=24$ and the {\it Roman} mission at $m_1=26$ over a smaller area, but extending to K band shown in the last column.
\textcolor{black}{When CMB dipole is entirely kinematic the CIB/IGL dipole components would be $S_i=(d_\nu/I_\nu){n}_i$ in the direction $\hat{n}=(-0.07,-0.66,0.75)$. Its direction uncertainties were evaluated per \cite{Atrio-Barandela:2010}:  the error on $S_i$ is given by that 
on the mean, $N_{\rm tot}^{-1/2}$, weighted by the dispersion of the direction
cosines on the sky area $4\pi f_{\rm sky}$ covered by the survey,
i.e., $N_i=1/(N_{\rm tot}\langle n_i^2\rangle)^{1/2}$ with
$\langle n_i^2\rangle=(4\pi)^{-1}\int_{4\pi f_{\rm sky}} n_i^2d\Omega$. 
We adopt the configurations from Table \ref{tab:tab_ngals}. 
Fig.\ref{fig:cib-s2n} shows the resultant $S/N$ accumulation for the CIB dipole components in each band for the two missions.  (For the {\it Euclid} example we used the projected sky coverage from \cite{Laureijs:2020}. For {\it Roman} the area that will be observed has not been yet finalized. For our estimates we used an area of $\sim 1,700$ deg$^2$ bounded in R.A. and Dec by $10\le\alpha\lt 60$ and  $-50\le\delta\lt -10$. We call this area the Southern {\it Roman} patch. Since the satellite could be observing a larger area, we also considered the final area to include another symmetric Northern {\it Roman} patch, limited by $190\le\alpha\lt 240$ and  $10\le\delta\lt 50$.) The Z-component's $S/N$ is the largest because of the planned avoidance of Galactic and Ecliptic planes.}

\textcolor{black}{Generally,} the CIB/IGL dipole direction will be well determined in these configurations with errors decreasing $\propto N_{\rm tot}^{-1/2}$ for the configurations in Table \ref{tab:tab_ngals}. If the CIB dipole points in the CMB dipole's direction, the directional uncertainty would be:
\begin{equation}
\Delta \Theta(90\%,95\%,99\%)=(1.8^\circ,2^\circ,2.6^\circ) \left(\frac{N_{\rm tot}}{10^9}\right)^{-1/2}
\label{eq:direction}
\end{equation}


%
\section{Discussion}

The IGL can measure the CIB dipole with $S/N\gsim 50-100$, testing the kinematic 
origin of the CMB dipole to $\sim 1^\circ$. 
Eq \ref{eq:dndm} can be subdivided into narrow magnitude bins ($m_0\rightarrow m_1$) where $\eta$ does not vary significantly but can be evaluated with the available data.  \textcolor{black}{Since  the CIB dipole due to local motion should be independent of redshift,
probing it for sources within various redshift ranges to $z\simeq 3$ will help establish its kinematic origin}. 
If the primordial component turns out to be present,
it could be generated by isocurvature perturbations of pre-inflationary origin,
opening a window to probe the superhorizon structure of the Universe prior to inflation.


By construction, there would not be any zodi or ISM impacts on the IGL dipole. Any IGL dipole should fall  with $\lambda$ in accordance with Fig. \ref{fig:cib} while 
 the foregrounds at the near-IR wavelengths come from reflected light \citep[][]{Kelsall:1998,Arendt:1998} and rise steeply toward the visible, providing a consistency test. If there is a
  systematic in the photometric zeropoints that has a dipole
  component, this will give a false dipole signature.  Given the
  expected CIB dipole of $\simeq 0.5\%$, one needs these
  systematics at $\lsim0.1\%$.

Extinction corrections (ECs) are important in the application of this method as discussed by \cite{Secrest:2021} who found the effect small.
They are more important in visible than in the near-IR bands and the {\it Roman}'s K band is where extinction is smallest.
Scaling the E(B-V) reddening maps of \cite{Schlegel:1998} indicates that
without ECs the dipole amplitude may be comparable with the IGL dipole. 
A direct approach to reduce this contribution is to apply an extinction
correction to the photometry of the individual galaxies before calculating the
IGL \citep{Secrest:2021}. ECs must be
sufficiently accurate so any residual errors are small compared to the
expected signal. Models can fit reddening coefficients or colors
to a few percent \cite[e.g.][]{Scaramella:2021}.
\textcolor{black}{The analyses could be restiricted to low extinction areas}. The cosmic dipole can be determined 
with an error $\lsim 10^\circ$ (90\% c.l.) with some $10^8$ galaxies and the error is even 
smaller for other configurations in Table \ref{tab:tab_ngals}. Then, the analysis can be extended to larger
areas. All the measured dipoles, further probed by {\it Roman} at greater depths, should be consistent and point in the same direction within
the errors with the amplitude scaling as in Fig. \ref{fig:cib} 
for all the selected patches, bands and configurations.

Galactic stars-galaxies separation is necessary to
avoid a spurious dipole from stars, which can dominate source
counts at $m \lesssim$18--20 in high latitude fields \cite[e.g. ][]{Windhorst:2011,
Ashby:2013}. Simple solutions based on source size \cite[e.g.][]{Windhorst:2011} will suffice here, because the exclusion of all compact
sources will eliminate a Galactic stellar dipole, and it will not induce a
cosmic dipole if compact extragalactic sources are inadvertently excluded. The further advantage of the forthcoming space missions' sub-arcsec angular resolution is the ability to resolve the galaxies individually, robustly overcoming source-confusion. 

This bodes well for a high precision measurement of the CIB dipole necessary to resolve the kinematic nature of the CMB dipole. The primordial CMB dipole component will be estimated robustly from the measured IGL dipole using Fig. \ref{fig:cib} to transform to velocity then converted into the kinematic CMB component to be subtracted from the CMB dipole of highly precise amplitude and direction.

\section*{Acknowledgements}

We particularly thank Rick Arendt for contributions to understanding the role of foregrounds and other systematics and Kari Helgason for the HRK reconstruction used here. 
FAB acknowledges Grants PGC2018-096038-B-I00 (MINECO and FEDER, ``A way of making Europe") and SA083P17 from the Junta de Castilla y Le\'on.

\section*{Data Availability}

No proprietary data were used here. The public data presented will be shared on reasonable request to the lead author (AK).


\label{lastpage}
\end{document}